\documentclass[twocolumn,showpacs]{revtex4}
\usepackage{graphics,epsfig,graphicx}
\usepackage{color}

\def\maj#1{\ifmmode\mbox{\usefont{U}{msb}{m}{n}#1}\else{\usefont{U}{msb}{m}{n}#1}\fi}
\def\v#1{\mathbf{#1}}

\begin{document}

\title{\textbf{BCS ansatz for superconductivity in the light of the Bogoliubov approach and the
Richardson-Gaudin exact wave function}}
\author{ M. Combescot$^{(1)}$, W. V. Pogosov$^{(1,2)}$, and O. Betbeder-Matibet$^{(1)}$\\
$^{(1)}$\small {\textit{Institut des NanoSciences de Paris,}}\\
\small {\textit{Universit\'e Pierre et Marie Curie, CNRS,}}\\
\small {\textit{4 Place Jussieu, 75252 Paris Cedex 05}}\\
$^{(2)}$\small{\textit{Institute for Theoretical and Applied Electrodynamics,}}\\
\small{\textit{Russian Academy of Sciences, Izhorskaya 13, 125412 Moscow, Russia}}}

\begin{abstract}
The Bogoliubov approach to superconductivity provides a strong
mathematical support to the wave function ansatz proposed by
Bardeen, Cooper and Schrieffer (BCS). Indeed, this ansatz --- with all
pairs condensed into the same state --- corresponds to the
ground state of the Bogoliubov Hamiltonian. Yet, this Hamiltonian
only is part of the BCS Hamiltonian. As a result, the BCS ansatz definitely
differs from the BCS Hamiltonian ground state. This can be directly shown
either through a perturbative approach starting from the
Bogoliubov Hamiltonian, or better by analytically solving the BCS
Schr\"{o}dinger equation along Richardson-Gaudin exact procedure.
Still, the BCS ansatz leads not only to the correct extensive part
of the ground state energy for an arbitrary number of pairs in the
energy layer where the potential acts --- as recently obtained by
solving Richardson-Gaudin equations analytically --- but also to a few
other physical quantities such as the electron distribution, as here shown. The
present work also considers arbitrary filling of the potential
layer and evidences the existence of a super dilute and a super dense
regime of pairs, with a gap \emph{different} from the usual gap.
These regimes constitute the lower and upper limits of density-induced BEC-BCS cross-over in Cooper pair systems.
\end{abstract}


\date{\today}

\maketitle

\section{Introduction}

A centenary ago, Kammerling
Onnes \cite{Onnes} discovered superconductivity which is one of the most
fascinating phenomena of Solid State Physics: indeed, it is so at odd
from usual understanding that half a century has been necessary to reach some
acceptable microscopic picture of the phenomenon. The first clue came from understanding that, in spite of their Coulomb repulsion, two electrons
can attract each other via the ion motion \cite{Frol}. Although very
small, the resulting effective attraction can produce
a two-electron bound state when acting in an energy range having a finite density
of states, as shown by Leon Cooper \cite{Cooper}. The next step was to note that fermion pairs
being boson-like particles, two-electron bound states should
condense into a collective state quite different from a ``normal''
electron gas. Although this
condensation was claimed to be totally different from Bose-Einstein condensation for
elementary bosons \cite{Schrieffer}, the wave function ansatz used in
the Bardeen-Cooper-Schrieffer (BCS) theory of superconductivity
\cite{BCS} yet is the grand canonical ensemble version of all pairs
condensed into the same state --- as seen more in detail below. To support
this difference, it is however clear that a state reading as the
product of two linear combinations of free electron pairs is
different from two individual products because, due to the Pauli
exclusion principle, one free fermion pair state is
``missing in the second pair''. Since this Pauli blocking effect increases
with pair number, the state reading as a product of $N$ identical
linear combinations of free electron pairs ends by being very different
from $N$ individual products, as it would be for $N$ elementary bosons in a
Bose-Einstein condensate.

Nevertheless, even if more and more free pair states are missing when the number of correlated pair states increases --- through what we called ``moth-eaten effect'' \cite{JETPLett} in the exciton context --- it is yet of importance to know if the picture of superconductivity resulting from a product of \emph{identical} electron pairs still is an acceptable picture of the phenomenon. This picture comes from an ansatz which, as formidable support, relies on the fact that it leads to easy calculations with results in agreement with experiments. Nevertheless, it is well known that wave functions are quite subtle quantities, different ones possibly giving same values for quantities as averaged as the energy.

The fact that Cooper pairs are not elementary but composite bosons
pushes us to question the widely spread idea that these fermion
pairs form a condensate with all pairs in the same linear
combination of free-pair states, as in the case of elementary
boson condensate, even if, clearly, some collective effect takes
place in this condensation, as evidenced by the fact that
the linear combination in the BCS ansatz is definitely
different from the one of a single pair found by
Cooper\cite{Cooper}. This question becomes even more relevant when
considering the work done by Richardson\cite{Rich} and by
Gaudin\cite{Gaudin} a few years after the BCS milestone paper on
superconductivityÊ\cite{BCS}. Indeed, these authors succeeded to
write the exact form of the $N$-pair eigenstates for the
Hamiltonian considered by Bardeen, Cooper and Schrieffer\cite{BCS}
--- which turns out to be one of the very few exactly solvable
models. This exact $N$-pair wave function reads as
\begin{equation}
B^\dag(R_1)\cdots B^\dag(R_N)|F_0\rangle\ ,
\end{equation}
where $B^\dag(R_i)$ is the following linear combination of up and down spin electrons,
\begin{equation}
B^\dag(R_i)=\sum_{\v k}\frac{\omega_{\v k}}{2\epsilon_{\v k}-R_i}a_{\v k\uparrow}^\dag a_{-\v k\downarrow}^\dag\ .
\end{equation}
$\omega_{\v k}$ is equal to 1 in the energy layer where the potential acts --- later on called ``potential layer''. $|F_0\rangle$ is the frozen core Fermi sea, made of electrons which do not feel the BCS potential, i.e., electrons for which $\omega_{\v k}=0$. The $R_i$'s are solution of $N$ coupled equations
\begin{equation}
1=V\sum_{\v k}\frac{\omega_{\v k}}{2\epsilon_{\v k}-R_i}+\sum_{j\neq i}\frac{2V}{R_i-R_j}\ .
\end{equation}
In view of the second term of the above equation which can be traced back to the Pauli exclusion principle \cite{Guojun}, the $R_i$'s must be all different. As a result, the exact wave function given in Eq.(1) evidently differs from $N$ pairs in the same linear combination, as in the BCS ansatz.

Over the last half century \cite{Duk}, these Richarson-Gaudin
equations stayed analytically unsolved for arbitrary pair number
$N$ and arbitrary potential strength $V$. In order to better
understand the effect of Pauli blocking between composite bosons
--- which imposes to consider a fixed number of pairs feeling the
potential and thus amounts to stay in the canonical ensemble ---
we tackled these Richardson-Gaudin equations again quite recently
\cite{JETPLett,PRB,EPJB}. In contrast to traditional BCS theory
of superconductivity which corresponds to fill half of the potential layer, the Richardson-Gaudin procedure allows us
to consider arbitrary fillings of this layer. We first
succeeded to get the $R_i$'s solution of  these Richardson's
equations in the dilute limit of pairs \cite{JETPLett}, and then
found a way \cite{Crouzeix} to reach their sum which is the
$N$-pair ground state energy $E_N=\sum_i R_i$, without having to
calculate the $R_i$'s individually (see also Ref. \cite{Pogosov}).
We found that $E_N$ takes a nicely compact $N$-dependence,
\begin{equation}
E_N=NE_1+\frac{N(N-1)}{\rho}\,\frac{1+\sigma}{1-\sigma}\ ,
\end{equation}
within underextensive terms in $(N/\rho)^n$ independent of sample volume and thus negligible in front of the two terms of Eq.(4). $\rho$ is the density of states taken as constant in the potential layer. $E_1=2\epsilon_{F_0}-2\Omega\sigma/(1-\sigma)$ with $\sigma=\exp(-2/\rho V)$ is the single pair energy found by Cooper, $\epsilon_{F_0}$ being the Fermi energy of the frozen core electrons, i.e., electrons which do not feel the potential. $\Omega$ is the potential layer extension, of the order of twice a phonon energy.

It is remarkable to note that this expression of the $N$-pair energy fully agrees not only with Cooper's result for $N=1$, but also with
the ground state energy obtained by Bardeen, Cooper and Schrieffer in the dense regime,
through a minimization of the Hamiltonian mean value calculated
within the wave function ansatz\cite{BCS} --- which fundamentally corresponds to
take the $N$ pairs all condensed into the \emph{same} linear
combination, while the exact wave function given in Eq.(1) definitely corresponds to
different linear combinations. Indeed, the BCS ground state energy
is known to read in the weak coupling limit, $\sigma\simeq 0$, as
\begin{equation}
E_{N_{BCS}}\simeq\frac{1}{2}\rho\,\Delta^2\ ,
\end{equation}
where, in this limit, the gap $\Delta$ scales as $\Omega\sqrt{\sigma}$. This result fully agrees with Eq.(4) in the case of half filling, i.e., a pair number equal to $N_{BCS}=\rho\Omega/2$, which corresponds to a potential extending symmetrically over a phonon energy scale, on both sides of the normal electron Fermi sea. Even if cases of wave functions very different from the exact one while giving the correct energy, have been reported, the fact that the exact $N$-pair energy does support the BCS wave function ansatz still is quite puzzling because, in this ansatz, the $N$ pairs are condensed all into the \emph{same} state while in the exact wave function, they all are in different states due Pauli blocking between the fermionic components of these composite bosons.

Actually, there is another approach to superconductivity different from the Richardson-Gaudin procedure which allows reaching some microscopic understanding of this ansatz, while evidencing that this ansatz formally differs from the exact wave function. It is based on the Bogoliubov approach to superconductivity. This approach allows an exact diagonalization of a \emph{part} of the original BCS Hamiltonian extended to the grand canonical ensemble. We here show that the ground state of this partial Hamiltonian corresponds to the BCS ansatz. Contributions to the ground state energy coming from the remaining part of the BCS Hamiltonian can be made negligible  in the thermodynamical limit, i.e., when the change from canonical to grand canonical ensemble is expected to be small. Nevertheless, this remaining part of the BCS Hamiltonian still changes the ground state itself, which thus formally differs from the BCS ansatz with all pairs condensed into the same state.

Such a conclusion \emph{on the ground state energy} has already been
reached by Boboliubov \cite{Bogoliubov} and also by Bardeen and
Rickayzen \cite{Bardeen} using single-particle Green functions, in
a form given by Galitskii \cite{Galitskii}. We also wish to
mention Mattis and Lieb's work  \cite{Lieb}, which
supports the results of BCS theory obtained within the BCS ansatz.

Although the ground state energy for sure is a major physical quantity, which explains why previous works concentrated on showing the validity of the BCS result for this quantity, the validity of the BCS ansatz for the ground state is of fundamental importance because this ground state ansatz is at the origin of our common microscopic understanding of superconductivity, ``with all pairs condensed into the same state''. The purpose of the present work is to concentrate on this ground state, in order to better understand why, even if formally different from the exact form, the BCS ansatz can still catch the correct energy as well as a few other correlation functions, such as the electron distribution.

The present paper is organized as follows.

In section II, we briefly recall the BCS model Hamiltonian for
superconductivity and the ground state ansatz proposed by Bardeen,
Cooper and Schrieffer, with particular emphazis on its projection to the $N$-pair subspace.

In section III, we propose a very simple presentation of the
Bogoliubov approach to the BCS problem based on splitting
the BCS Hamiltonian $H_{BCS}$ extended to the grand canonical ensemble, as $\hat{H}_{BCS}=\hat{H}_B+W$, these two parts depending on a set of
scalars which, at this stage, are taken as arbitrary. We then show how to easily diagonalize the "Bogoliubov Hamiltonian" $\hat{H}_B$.

In section IV, we discuss some properties of the "Bogoliubov
Hamiltonian" eigenstates. In particular, we show that the form of $\hat{H}_B$ ground state just corresponds
to the BCS ansatz. This gives a
strong mathematical support to this ansatz in spite of its obvious
difference with the exact form of the BCS ground state obtained by Richardson and by Gaudin.

In section V, we introduce the difference $W=\hat{H}_{BCS}-\hat{H}_B$ between the BCS Hamiltonian and the Bogoliubov Hamiltonian as a perturbation, and we
determine the set of scalars introduced in the Bogoliubov procedure which makes the $W$
contributions to the $\hat{H}_{BCS}$ ground state energy underextensive.

In section VI, we consider the mean values of a few physical operators in the $\hat{H}_B$ ground state and in the $\hat{H}_{BCS}$ ground state calculated through perturbation theory in $W$. We show that the scalars introduced in the Bogoliubov procedure which leads to the correct value for the ground state energy, also give the correct value for a few relevant quantities such as the electron distribution.

In section VII, we come back to the ground state of the
Bogoliubov Hamiltonian $\hat{H}_B$ and consider not only the
standard BCS half-filling configuration, but also an arbitrary
filling of the potential layer. Although these arbitrary fillings
are difficult to achieve experimentally, their analysis allows a
deeper understanding of Cooper-paired states by bridging two
textbook problems, namely, the single Cooper pair problem and the
BCS model for superconductivity. It is worth noting,
as argued in Ref.~\cite{Geyer}, that these arbitrary fillings
can be relevant in some semiconductor situations (see also
Ref.~\cite{Eagles}).

In Section VIII, we consider the $\hat{H}_B$ excited states and
derive the energy gap for an arbitrary filling of the potential
layer. This study reveals the existence of an extremely dilute and
an extremely dense regime of pairs with a gap different
from the usual gap. These two regimes have to be considered in a complete description of the density induced BEC-BCS cross-over in
Cooper pair systems along the line proposed by Eagles\cite{Eagles} and by
Leggett\cite{Tonycross}.

In the last section, we present some concluding comments.

\section{ BCS wave function ansatz}

We consider a system made of $N$ pairs of different fermions $\alpha$ and $%
\beta$. In the case of BCS superconductivity, these two fermions are the up and down
spin electrons. For fermions with same mass, the Hamiltonian free part reads in terms of their creation operators $a_{\mathbf{k}}^\dag$
and $b_{\mathbf{k}}^\dag$ as
\begin{equation}
H_0=\sum_{\mathbf{k}}\epsilon_{\mathbf{k}}(a_{\mathbf{k}}^\dag a_{\mathbf{k}%
}+b_{\mathbf{k}}^\dag b_{\mathbf{k}})\ .
\end{equation}
In usual BCS pairing, fermion-fermion attraction is reduced to
zero-momentum pair processes. So, the total Hamiltonian reads as $H_{BCS}=H_0+V_{BCS}$
with
\begin{equation}
V_{BCS}=-\sum_{\mathbf{k}^{\prime},\mathbf{k}}V_{\mathbf{k}^{\prime}\mathbf{k%
}}a_{\mathbf{k}^{\prime}}^\dag b_{-\mathbf{k}^{\prime}}^\dag b_{-\mathbf{k}%
}a_{\mathbf{k}}\ .
\end{equation}
In order to end with an analytically solvable problem, we will ultimately take these $V_{%
\mathbf{k}^{\prime}\mathbf{k}}$ scatterings in a
separable form $V\omega_{\v k'}\omega_{\v k}$.

We want to determine the ground state of these $N$ pairs. In order to have
the lowest possible energy, these pairs must enjoy the attractive potential as much as
possible; so, pairs from which the ground state is formed must have a zero momentum, i.e., a creation operator reading as
\begin{equation}
B_{\mathbf{k}}^\dag=a_{\mathbf{k}}^\dag b_{-\mathbf{k}}^\dag\ .
\end{equation}
We thus expect the ground state for $N$ pairs to read as
\begin{equation}
|\psi_N^{(0)}\rangle=\sum_{\mathbf{k}_1,\ldots,\mathbf{k}_N}F(\mathbf{k}%
_1,\ldots,\mathbf{k}_N)B_{\mathbf{k}_1}^\dag\cdots B_{\mathbf{k}%
_N}^\dag|F_0\rangle\ ,
\end{equation}
with $(\mathbf{k}_1,\cdots,\mathbf{k}_N)$ all different due to the
Pauli exclusion principle.

To exactly handle this Pauli blocking is definitely difficult. A
smart way to overcome this difficulty is to turn to the grand
canonical ensemble, with a pair number not fixed as proposed by
Bardeen, Cooper and Schrieffer through an ansatz constructed on the
idea that, electron pairs being boson-like particles, they are
likely to condense all into the same state. By writing this correlated state as
\begin{equation}
B^\dag=\sum_{\mathbf{k}}\varphi_{\mathbf{k}}B_{\mathbf{k}}^\dag\ .
\end{equation}
Bardeen, Cooper and Schrieffer proposed a ground state ansatz in the grand canonical ensemble reading as
\begin{equation}
|\phi_{BCS}\rangle=\sum_{N=1}^{+\infty}\frac{1}{N!}B^{\dag N}|F_0\rangle\ .
\end{equation}
Note that the $B^{\dag N}$ prefactor, chosen as $1/N!$ to possibly
perform the sum over $N$ easily, is unimportant in the
thermodynamical limit, i.e., when the grand-canonical approach is
valid, because the $N$ distribution is known to be very much peaked on the $N$
mean value - as possible to explicitly show (see
Ref.~\cite{Mon}). The state $|\phi_{BCS}\rangle$ then takes a compact form
\begin{eqnarray}
|\phi_{BCS}\rangle =e^{B^\dag}|F_0\rangle=\prod_{\mathbf{k}}e^{\varphi_{%
\mathbf{k}}B_{\mathbf{k}}^\dag}|F_0\rangle  \nonumber \\
=\prod_{\mathbf{k}}(1+\varphi_{\mathbf{k}}B_{\mathbf{k}}^\dag)|F_0\rangle,
\hspace{0.5cm}
\end{eqnarray}
since $B_{\mathbf{k}}^{\dag 2}|F_0\rangle=0$ due to Pauli blocking.
By writing $\varphi_{\mathbf{k}}$ as
$v_{\mathbf{k}}/u_{\mathbf{k}}$, we readily recover the usual form
of the BCS wave function ansatz, namely,
 \begin{equation}
|\psi_{BCS}\rangle=\prod_{\mathbf{k}}(u_{\mathbf{k}}+v_{\mathbf{k}}B_{%
\mathbf{k}}^\dag)|F_0\rangle\ ,
\end{equation}
within an irrelevant prefactor.
The $(u_{\v k},v_{\v k})$ coefficients, enforced to fulfill $|u_{\mathbf{k}}|^2+|v_{\mathbf{k}}|^2=1$ in order to have $%
\langle\psi_{BCS}|\psi_{BCS}\rangle=1$ are ultimately determined by a variational procedure (see for example Ref.  \cite{Fetter}), through the
minimization of the Hamiltonian mean-value $\langle\psi_{BCS}|\hat{H}_{BCS}|\psi_{BCS}%
\rangle$, the BCS Hamiltonan in the grand canonical ensemble reading as
\begin{equation}
\hat{H}_{BCS}=H_{BCS}-\mu\hat{N}=\hat{H}_0+V_{BCS}\ ,
\end{equation}
where $\hat{N}$ is the number operator for fermions $\alpha$ and $\beta$, namely,
\begin{equation}
\hat{N}=\sum_{\mathbf{k}}(a_{\mathbf{k}}^\dag a_{\mathbf{k}}+b_{\mathbf{k}%
}^\dag b_{\mathbf{k}})\ .
\end{equation}
As a result, the one-body part of the $\hat{H}_{BCS}$ Hamiltonian reads as
\begin{equation}
\hat{H}_0=\sum_{\v k}\xi_{\v k}\left(a_{\v k}^\dag a_{\v k}+b_{\v k}^\dag b_{\v k}\right)\ ,
\end{equation}
with $\xi_{\v k}=\epsilon_{\v k}-\mu$.

We wish to note that, although the parameters $u_{\v k}$ and $v_{\v k}$ are quite popular in all textbooks dealing with superconductivity, the key parameters of this problem are not ($u_{\v k}$, $v_{\v k}$) but their ratio $\varphi_{\v k}=v_{\v k}/u_{\v k}$. Moreover, from a practical point of view, minimization with respect to $(u_{\v k},v_{\v k})$, as down in the standard BCS procedure, implies to handle four quantities, namely, two modulus and two phases, while in the minimization with respect to $\varphi_{\v k}$, we only have to determine one modulus and one phase. To our opinion, the introduction of $(u_{\v k},v_{\v k})$ is useless, as also seen in a transparent way using the Bogoliubov approach to the BCS problem.

\section{Bogoliubov approach}

\subsection{Bogoliubov Hamiltonian}

It is possible to give a precise mathematical meaning to the ground state ansatz proposed by Bardeen, Cooper and Schrieffer, by using the
Bogoliubov approach to BCS superconductivity. A very simple way to present this
approach is to introduce a set of yet arbitrary complex scalars $z_{\mathbf{k%
}}$ and to split the BCS potential given in Eq.(7) as
\begin{equation}
V_{BCS}=\mathcal{E}+V_B+W\ .
\end{equation}
The two-body character of the BCS interaction is concentrated into the operator $W$ given by
\begin{equation}
W=-\sum_{\mathbf{k}^{\prime},\mathbf{k}}V_{\mathbf{k}^{\prime}\mathbf{k}}(B_{%
\mathbf{k}^{\prime}}^\dag-z_{\mathbf{k}^{\prime}}^\ast)(B_{\mathbf{k}}-z_{%
\mathbf{k}})\ .
\end{equation}
The $z_{\mathbf{k}}$ scalars are ultimately chosen to make the $W$
contributions to the ground state energy negligible in the large sample limit.

By extracting $W$ from $V_{BCS}$, we are left with a $V_B$ potential quadratic in electron operators, which is the goal of the procedure. It reads
\begin{equation}
V_B=-\sum_{\mathbf{k}^{\prime}}\Delta_{\mathbf{k}^{\prime}}B_{\mathbf{k}
^{\prime}}^\dag +\mathrm{h.c.}\ ,
\end{equation}
where the prefactor $\Delta_{\mathbf{k}^{\prime}}$ depends on $V_{\mathbf{k}^{\prime}\mathbf{k}}$
and on the arbitrary scalars $z_{\mathbf{k}}$ as
\begin{equation}
\Delta_{\mathbf{k}^{\prime}}=\sum_{\mathbf{k}}V_{\mathbf{k}^{\prime}\mathbf{k
}}\,z_{\mathbf{k}}\ .
\end{equation}
This gives the scalar $\mathcal{E}$ in Eq.(17) as
\begin{equation}
\mathcal{E}=\sum_{\mathbf{k}^{\prime},\mathbf{k}}V_{\mathbf{k}^{\prime}
\mathbf{k}}\,z_{\mathbf{k}^{\prime}}^\ast z_{\mathbf{k}}\ .
\end{equation}
This scalar is real since $V_{\mathbf{k}^{\prime}\mathbf{k}}=V_{\mathbf{k}
\mathbf{k}^{\prime}}^\ast$ as required from $V_{BCS}=V_{BCS}^\dag$.

While $V_{BCS}$ conserves the particle number, $V_B$ and $
W$ do not conserve this number separately. In order to possibly treat $W$ as a
perturbation independently from $V_B$, it is then mandatory to turn to the
grand canonical ensemble. This leads us to turn from $H_{BCS}$ to the Hamiltonian $\hat{H}_{BCS}$
in this grand canonical ensemble. Using Eq.(17), $\hat{H}_{BCS}$ then splits as
\begin{equation}
\hat{H}_{BCS}=\hat{H}_B+W\ ,
\end{equation}
where $W$ is given by Eq.(18) while $\hat{H}_B$ given by
\begin{eqnarray}
\hat{H}_B=\mathcal{E}+\sum_{\mathbf{k}}\left\{\xi_{\mathbf{k}}(a_{\mathbf{k}}^\dag
a_{\mathbf{k}}+b_{-\mathbf{k}}^\dag b_{-\mathbf{k}})\right.  \nonumber \\
\left.-(\Delta_{\mathbf{k}}B_{\mathbf{k}}^\dag+\mathrm{h.c.})\right\}\
\end{eqnarray}
is what we are going to call ``Bogoliubov Hamiltonian''

\subsection{Diagonalization of the Bogoliubov Hamiltonian}

The $\hat{H}_B$ Hamiltonian, quadratic in fermion operators, is easy to diagonalize. For that, we introduce new operators defined as
\begin{eqnarray}
\tilde{a}_{\mathbf{k}}^\dag&=&x_{\mathbf{k}}a_{\mathbf{k}}^\dag+y_{\mathbf{k}
}b_{-\mathbf{k}}  \nonumber \\
\tilde{b}_{-\mathbf{k}}^\dag&=&x_{\mathbf{k}}^{\prime}b_{-\mathbf{k}
}^\dag+y_{\mathbf{k}}^{\prime}a_{\mathbf{k}}\ .
\end{eqnarray}
We first enforce these new operators to have the same anticommutation relations as the
original fermion operators $a_{\mathbf{k}}^\dag$ and $b_{-\mathbf{k}}^\dag$. This imposes the prefactors in Eq.(24) to be such that
\begin{equation}
\left[\tilde{a}_{\mathbf{k}},\tilde{a}_{\mathbf{k}}^\dag\right]_+=1=|x_{%
\mathbf{k}}|^2+|y_{\mathbf{k}}|^2\ ,
\end{equation}
with a similar relation for $(x'_{\v k},y'_{\v k})$, while
\begin{equation}
\left[\tilde{a}_{\mathbf{k}}^\dag,\tilde{b}_{-\mathbf{k}}^\dag\right]_+=0=y_{%
\mathbf{k}}\,x_{\mathbf{k}}^{\prime}+x_{\mathbf{k}}\,y_{\mathbf{k}%
}^{\prime}\ ,
\end{equation}
the three other anticommutators,
$\left[\tilde{a}_{\mathbf{k}}^\dag,\tilde{a}_{\mathbf{k}}^\dag\right]_+$,
$\left[\tilde{b}_{-\mathbf{k}}^\dag,\tilde{b}_{-\mathbf{k}}^\dag\right]_+$ and
$\left[\tilde{a}_{\mathbf{k}},\tilde{b}_{-\mathbf{k}}^\dag\right]_+$
being automatically equal to zero due to the anticommutation relations existing between the $a_{\v k}^\dag$ and $b_{-\v k}^\dag$ operators.

Equation (26) leads us to introduce $f_{\v k}$ defined as
\begin{equation}
f_{\v k}=\frac{y_{\mathbf{k}}}{x_{\mathbf{k}}}=-\frac{y_{\mathbf{k}}^{\prime}}{x_{%
\mathbf{k}}^{\prime}}\ .
\end{equation}
When inserted into Eq.(25) for $(x_{\v k},y_{\v k})$ and the similar equation for $(x'_{\v k},y'_{\v k})$, we readily get
\begin{equation}
|x_{\mathbf{k}}|^2=|x_{\mathbf{k}}^{\prime}|^2=\frac{1}{1+|f_{\v k}|^2}\ .
\end{equation}

Next, we note that the product of Bogoliubov operators defined in Eq.(24) reads as
\begin{eqnarray}
\tilde{a}_{\mathbf{k}}^\dag\tilde{a}_{\mathbf{k}}=|y_{\v k}|^2+|x_{\mathbf{k}}|^2a_{%
\mathbf{k}}^\dag a_{\mathbf{k}}-|y_{\mathbf{k}}|^2b_{-\mathbf{k}}^\dag b_{-\mathbf{%
k}}  \nonumber \\
+\Big{(}y_{\mathbf{k}}x_{\mathbf{k}}^\ast B_{\mathbf{k}}+\mathrm{h.c.}\Big{)}\ ,
\end{eqnarray}
since $b_{-\mathbf{k}}b_{-\mathbf{k}}^\dag=1-b_{-\mathbf{k}}^\dag b_{-
\mathbf{k}}$, with a similar expression for the product $\tilde{b}_{-\mathbf{k}}^\dag\tilde{b}_{-\mathbf{k}}$.
So, using Eq.(28), we get the sum of these two products as
\begin{eqnarray}
\tilde{a}_{\mathbf{k}}^\dag\tilde{a}_{\mathbf{k}}+\tilde{b}_{-\mathbf{k}
}^\dag\tilde{b}_{-\mathbf{k}}=\left(1+|f_{\v k}|^2\right)^{-1}\hspace{2cm}\nonumber\\
\times\Big{\{}2|f_{\v k}|^2
+\left(1-|f_{\v k}|^2\right)\left(a_{\v k}^\dag a_{\v k}+b_{-\v k}^\dag b_{-\v k}\right)\Big{.}\nonumber\\
\Big{.}+2\left(f_{\v k}B_{\v k}+\mathrm{h.c.}\right)\Big{\}}\ .
\end{eqnarray}
It is then easy to see that we can identify the Bogoliubov Hamiltonian defined in Eq.(23) with
\begin{equation}
\hat{H}_B=\hat{\mathcal{E}}_B+\sum_{\mathbf{k}}\tilde{E}_{\mathbf{k}}(\tilde{a}_{\mathbf{k}}^\dag%
\tilde{a}_{\mathbf{k}}+\tilde{b}_{-\mathbf{k}}^\dag\tilde{b}_{-\mathbf{k}})\ ,
\end{equation}
provided that we set
\begin{eqnarray}
\mathcal{E}&=&\hat{ \mathcal{E}}_B+\sum_{\v k}\tilde{E}_{\v k}\frac{2|f_{\v k}|^2}{1+|f_{\v k}|^2}\ ,\\
\xi_{\v k}&=&\tilde{E}_{\v k}\frac{1-|f_{\v k}|^2}{1+|f_{\v k}|^2}\ ,\\
-\Delta_{\v k}&=&\tilde{E}_{\v k}\frac{2f_{\v k}^\ast}{1+|f_{\v k}|^2}\ .
\end{eqnarray}

The second condition readily gives
\begin{equation}
|f_{\v k}|^2=\frac{1-\xi_{\v k}/\tilde{E}_{\v k}}{1+\xi_{\v k}/\tilde{E}_{\v k}}\ ,
\end{equation}
which imposes $\tilde{E}_{\v k}$ real. From the third condition, we then get
$\tilde{E}_{\v k}^2=\xi_{\v k}^2+|\Delta_{\v k}|^2$
which is fulfilled for $\tilde{E}_{\v k}=\pm\sqrt{\xi_{\v k}^2+|\Delta_{\v k}|^2}$. The choice of a plus sign follows from noting that, if $\tilde{E}_{\v k}$ were taken negative, the $\tilde{H}_B$ ground state would have an infinite negative energy, which is unphysical. So, the physically relevant $\tilde{E}_{\v k}$ must be
\begin{equation}
\tilde{E}_{\v k}=\sqrt{\xi_{\v k}^2+|\Delta_{\v k}|^2}\ .
\end{equation}

This plus sign also follows from self-consistency at the ``gap equation'' level, as shown below.
Before going further, we wish to stress that this sign problem is linked to the fact that the $V_{BCS}=0$ limit for which all $\Delta_{\v k}$'s cancel, is highly singular. Indeed, Eq.(34) then gives $f_{\v k}=0$; so, $\xi_{\v k}=\tilde{E}_{\v k}$ due to Eq.(33) and $|x_{\v k}|=|x'_{\v k}|=1$ due to Eq.(28). As a result, we find $\tilde{a}_{\v k}^\dag=a_{\v k}^\dag$ and $\tilde{b}_{-\v k}^\dag=b_{-\v k}^\dag$ within an irrelevant phase factor. Everything is then consistent with $\hat{H}_B$ for $V_{BCS}=0$ being given by Eq.(31). However, $\xi_{\v k}=\tilde{E}_{\v k}$ for all $\v k$'s is inconsistent with Eq.(36), since for $\Delta_{\v k}=0$, this choice would lead to $\tilde{E}_{\v k}=|\xi_{\v k}|$, not $\tilde{E}_{\v k}=\xi_{\v k}$. So, $\tilde{E}_{\v k}$ and $\xi_{\v k}$ would not be equal but opposite for half of the $\v k$ values, in the case of a chemical potential $\mu$ taken in the middle of the potential layer, as in the BCS configuration. We can also note that
Eq.(28) with $|f_{\v k}|^2$ given in Eq.(35), leads to
\begin{equation}
|x_{\v k}|^2=1-|y_{\v k}|^2=\frac{1}{2}\left(1+\frac{\xi_{\v k}}{\tilde{E}_{\v k}}\right)\ .
\end{equation}
Equation (36) then gives, for a vanishingly small potential, $|x_{\v k}|\simeq 0$ for $\xi_{\v k}$ negative and $|x_{\v k}|\simeq 1$ for $\xi_{\v k}$ positive. As a result, $\tilde{a}_{\v k}^\dag$ must be identified with $a_{-\v k}$ for $\epsilon_{\v k}$ below the chemical potential $\mu$ and with $a_{\v k}^\dag$ above $\mu$. By contrast, the solution $\tilde{E}_{\v k}=\xi_{\v k}$ obtained for $V_{BCS}$ exactly equal to zero, gives $\tilde{a}_{\v k}^\dag=a_{\v k}^\dag$ for all $\v k$'s.

If we now turn to the scalar $\hat{\mathcal{E}}_B$ in the diagonal form of the ``Bogoliubov Hamiltonian'' given in Eq.(31), we find, using Eqs.(21,32), that it reads
\begin{equation}
\hat{\mathcal{E}}_B=\sum_{\mathbf{k}^{\prime},\mathbf{k}}V_{\mathbf{k}^{\prime}%
\mathbf{k}}\,z_{\mathbf{k}^{\prime}}^\ast z_{\mathbf{k}}+\sum_{\mathbf{k}}(\xi_{%
\mathbf{k}}-\tilde{E}_{\mathbf{k}})\ .
\end{equation}

\section{Eigenstates of the Bogoliubov Hamiltonian}

Equation (31) readily shows that the ground state energy of the Bogoliubov Hamiltonian $\hat{H}_B$ is equal to $\hat{\mathcal{E
}}_B$ while the excited states read as products of operators $\tilde{a}_{%
\mathbf{k}}^\dag$ and $\tilde{b}_{-\mathbf{k}}^\dag$ acting on the $\hat{H}_B$
ground state, their energy being a few $\tilde{E}_{\v k}$ above $\hat{\mathcal{E}}_B$. Moreover, Eq.(36) already shows that these excitations are associated with an energy gap $|\Delta_{\v k}|$.

\subsection{$\hat{H}_B$ ground state}

Let us note as $|0_B\rangle$ the $\hat{H}_B$ ground state. This state is, by construction, such that
\begin{equation}
\tilde{a}_{\v k}|0_B\rangle=0=\tilde{b}_{-\v k}|0_B\rangle\ .
\end{equation}
We are going to show that this ground state reads as the BCS ansatz, namely,
\begin{equation}
|0_B\rangle=\prod_{\v k}\big{(}1+g_{\v k}B_{\v k}^\dag\big{)}|F_0\rangle\ ,
\end{equation}
with $g_{\v k}$ related to the $f_{\v k}$ function introduced in Eq.(27) through
\begin{equation}
g_{\v k}=-f_{\v k}^{*}=\frac{y_{\v k}'^{*}}{x_{\v k}'^{*}}.
\end{equation}

To do so, we first note that the $\hat{H}_B$ ground state in the grand canonical ensemble is a priori made of zero, one, two, $\ldots$ free pair states $B_{\v k}^\dag$. Its most general form thus reads
\begin{eqnarray}
|0_B\rangle&=&\Big{[}1+\sum_{\v k}h_{\v k}B_{\v k}^\dag+\sum_{\v k,\v k'}h_{\v k\v k'}B_{\v k}^\dag B_{\v k'}^\dag+\cdots\Big{]}|F_0\rangle\nonumber\\
&\equiv&|0_B^{(0)}\rangle +|0_B^{(1)}\rangle+|0_B^{(2)}\rangle+\cdots\ ,
\end{eqnarray}
where $|0^{(n)}\rangle$ contains $n$ pair states.

Since Eqs.(24) and  Eq.(27) give the Bogoliubov operators $\tilde{a}_{\v k}$ and $\tilde{b}_{-\v k}$ as
\begin{eqnarray}
\tilde{a}_{\v k}&=&x_{\v k}^\ast\big{(}a_{\v k}+f_{\v k}^\ast b_{-\v k}^\dag\big{)}\nonumber\\
\tilde{b}_{-\v k}&=&x_{\v k}'^{\ast}\big{(}b_{-\v k}-f_{\v k}^\ast a_{\v k}^\dag\big{)}\ .
\end{eqnarray}
we readily find that $\tilde{a}_{\v k}$ and $\tilde{b}_{-\v k}$, acting on the zero-pair component $|0_B^{(0)}\rangle$ of the $|0_B\rangle$ ground state, give
\begin{eqnarray}
\tilde{a}_{\v p}|0_B^{(0)}\rangle&=&x_{\v p}^\ast f_{\v p}^\ast b_{-\v p}^\dag|F_0\rangle\nonumber\\
\tilde{b}_{-\v p}|0_B^{(0)}\rangle&=&-x_{\v p}'^\ast f_{\v p}^\ast a_{\v p}^\dag|F_0\rangle\ .
\end{eqnarray}
In the same way, when acting on the one-pair component $|0_B^{(1)}\rangle$ of $|0_B\rangle$, these operators give
\begin{eqnarray}
\tilde{a}_{\v p}|0_B^{(1)}\rangle&=&x_{\v p}^\ast\Big{[}h_{\v p}+f_{\v p}^\ast\sum_{\v k}h_{\v k}B_{\v k}^\dag\Big{]}b_{-\v p}^\dag|F_0\rangle\nonumber\\
\tilde{b}_{-\v p}|0_B^{(1)}\rangle&=&-x_{\v p}'^\ast\Big{[}h_{\v p}+f_{\v p}^\ast\sum_{\v k}h_{\v k}B_{\v k}^\dag\Big{]}a_{\v p}^\dag|F_0\rangle\ .
\end{eqnarray}
And so on \ldots for the higher-order components $|0_B^{(n)}\rangle$ of $|0_B\rangle$. By collecting all these terms, we end with
\begin{eqnarray}
\tilde{a}_{\v p}|0_B\rangle=x_{\v p}^\ast G_{\v p}b_{-\v p}^\dag|F_0\rangle\nonumber\\
\tilde{b}_{-\v p}|0_B\rangle=-x_{\v p}'^\ast G_{\v p}a_{\v p}^\dag|F_0\rangle\ ,
\end{eqnarray}
with the same $G_{\v p}$ in both equations, this $G_{\v p}$ being given by
\begin{eqnarray}
G_{\v p}=f_{\v p}^\ast+\Big{[}h_{\v p}+f_{\v p}^\ast\sum_{\v k}h_{\v k}B_{\v k}^\dag\Big{]}\hspace{2.5cm}\nonumber\\
+\left[\sum_{\v k}(h_{\v p\v k}+h_{\v k\v p})B_{\v k}^\dag+f_{\v p}^\ast\sum_{\v k,\v k'}h_{\v k\v k'}B_{\v k}^\dag B_{\v k'}^\dag\right]+\cdots
\end{eqnarray}

To go further, we project $\tilde{a}_{\v p}|0_B\rangle$ and $\tilde{b}_{-\v p}|0_B\rangle$ in Eq.(46) over the zero-pair state $|F_0\rangle$, the one-pair state $B_{\v p_1}^\dag|F_0\rangle$, the two-pair state $B_{\v p_1}^\dag B_{\v p_2}^\dag|F_0\rangle$, and so on $\ldots$, and we enforce these projections to be equal to zero due to Eq.(39). By noting that, for $B_{\v k}^\dag=a_{\v k}^\dag b_{-\v k}^\dag$, the scalar product of two two-pair states reduces to
\begin{eqnarray}
\langle F_0|B_{\v p_1} B_{\v p_2}  B_{\v k}^\dag B_{\v k'}^\dag|F_0\rangle=\hspace{3cm}\nonumber\\
\left(\delta_{\v p_1,\v k}\delta_{\v p_2,\v k'}+\delta_{\v p_1,\v k'}\delta_{\v p_2,\v k}\right)
(1-\delta_{\v p_1,\v p_2})\ ,
\end{eqnarray}
we find that these projections over $|F_0\rangle$, $B_{\v p_1}^\dag|F_0\rangle$, $B_{\v p_1}^\dag B_{\v p_2}^\dag |F_0\rangle$ respectively give
\begin{eqnarray}
h_{\v p}&=&-f_{\v p}^\ast\nonumber\\
h_{\v p\v p_1}+h_{\v p_1\v p}&=&-h_{\v p_1}f_{\v p}^\ast=h_{\v p}h_{\v p_1}\nonumber\\
h_{\v p\v p_1\v p_2}+(5\ \mathrm{perm})&=&h_{\v p}h_{\v p_1}h_{\v p_2}\ ,
\end{eqnarray}
and so on \ldots for the higher-order components of $|0_B\rangle$. When compared to the RHS of Eq.(40) in which each $\v k$ state is counted once only by construction, while in Eq.(42) the $B_{\v p_1}^\dag B_{\v p_2}^\dag|F_0\rangle$ state  appears for $(\v k=\v p_1,\v k'=\v p_2)$ and $(\v k=\v p_2,\v k'=\v p_1)$, the above set of equations shows that the expansion (42) of the $|0_B\rangle$ ground state is identical to the product given in Eq.(40), provided that we set
\begin{equation}
g_{\v p}=h_{\v p}=-f_{\v p}^\ast\ ,
\end{equation}
in agreement with Eq.(41).

\subsection{$\hat{H}_B$ excited states}

The Bogoliubov procedure is really nice because it also allows reaching the $\hat{H}_B$ excited states in a trivial way: the expression of the $\hat{H}_B$ Hamiltonian given in Eq.(31) readily shows that $\tilde{a}_{\v p}^\dag|0_B\rangle$ is $\hat{H}_B$ eigenstate with the energy $\hat{\mathcal{E}}_B+\tilde{E}_{\v p}$. This state physically corresponds to have the $B_{\v p}^\dag$ pair of the $|0_B\rangle$ ground state broken, with the $\tilde{b}_{-\v p}^\dag$ component removed. Indeed, by writing $|0_B\rangle$ in Eq.(40) as
\begin{equation}
|0_B\rangle=(1+g_{\v p}B_{\v p}^\dag)|0_{\v p}\rangle\ ,
\end{equation}
where $|0_{\v p}\rangle$ reads as $|0_B\rangle$ with the $\v p$ term in the $\v k$ product missing, we find, using Eqs.(24,27,28,41)
\begin{eqnarray}
\tilde{a}_{\v p}^\dag |0_B\rangle&=&(x_{\v p}a_{\v p}^\dag+y_{\v p}b_{-\v p})(1+g_{\v p}B_{\v p}^\dag)|0_{\v p}\rangle\nonumber\\
&=&(x_{\v p}-y_{\v p}g_{\v p})a_{\v p}^\dag|0_{\v p}\rangle\nonumber\\
&=&x_{\v p}(1+|f_{\v p}|^2)\,a_{\v p}^\dag|0_{\v p}\rangle\ .
\end{eqnarray}
In the same way, the excited state
\begin{equation}
\tilde{b}_{-\mathbf{p}}^\dag|0_B\rangle=x'_{\v p}(1+|f_{\v p}|^2)b_{-\mathbf{p}}^\dag|0_{\mathbf{p}%
}\rangle\
\end{equation}
has the $B_{\v p}^\dag$ pair of the $|0_B\rangle$ ground state broken, with the $a_{\v p}^\dag$ component removed.

\section{Eigenstates of the $\hat{H}_{BCS}$ Hamiltonian}

Through the above procedure, we have extracted the unpleasant
two-body part of the original BCS Hamiltonian, through $W$. This makes the $\hat{H}_B$
eigenstate spectrum straightforward to find. In order
for these eigenstates to have some connexion with the original BCS problem,
the ``perturbation'' $W$ has to bring negligible contributions. We have
some flexibility to reach this goal through the set of arbitrary scalars $z_{%
\mathbf{k}}$ introduced when rewriting the BCS potential as in Eqs.(17,18).

\subsection{Compact form of the perturbative expansion}

Before going further, let us first write the $\hat{H}_{BCS}$ ground state in terms of the $\hat{H}_B$ ground state in a compact form. We start with $(\hat{H}_B-\hat{\mathcal{E}}_B)|0_B\rangle=0$, which follows from Eq.(31) and we look for the $\hat{H}_{BCS}$ ground state,
\begin{equation}
\left(\hat{H}_{BCS}-\hat{\mathcal{E}}_{BCS}\right)|0_{BCS}\rangle=0\ ,
\end{equation}
in terms of the $\hat{H}_B$ ground state by using the identity
\begin{equation}
1=\frac{|0_B\rangle\langle0_B|}{\langle 0_B|0_B\rangle}+P_{\perp}\ ,
\end{equation}
which defines the projector $P_{\perp}$ over the subspace perpendicular to $|0_B\rangle$. This identity readily gives
\begin{equation}
|0_{BCS}\rangle=|0_B\rangle\,\frac{\langle 0_B|0_{BCS}\rangle}{\langle 0_B|0_B\rangle}+P_{\perp}|0_{BCS}\rangle\ .
\end{equation}

The next step is to calculate $P_{\perp}|0_{BCS}\rangle$. To do so,  we multiply Eq.(54) by $P_{\perp}$. This gives
\begin{equation}
0=P_{\perp}(\hat{H}_B-\hat{\mathcal{E}}_{BCS})|0_{BCS}\rangle+P_{\perp}W|0_{BCS}\rangle\ .
\end{equation}
We then inject Eq.(55) in front of the $|0_{BCS}\rangle$ state in the first term of the above equation. Since $P_{\perp}|0_B\rangle=0$, this leads to
\begin{equation}
P_{\perp}(\hat{\mathcal{E}}_{BCS}-\hat{H}_B)P_{\perp}|0_{BCS}\rangle=P_{\perp}W|0_{BCS}\rangle\ ,
\end{equation}
from which we get
\begin{equation}
P_{\perp}|0_{BCS}\rangle=P_{\perp}\frac{1}{\hat{\mathcal{E}}_{BCS}-\hat{H}_B}P_{\perp}W|0_{BCS}\rangle\ ,
\end{equation}
as easy to check by multiplying the above equation by $P_{\perp}(\hat{\mathcal{E}}_{BCS}-\hat{H}_B)$ and by writing $P_{\perp}$ as a sum of projections $|n_B\rangle\langle n_B|$ over all normalized $|n_B\rangle$ eigenstates of the $\hat{H}_B$ Hamiltonian except $|0_B\rangle$.

Equation (56) along with Eq.(59) then give
\begin{eqnarray}
|0_{BCS}\rangle=|0_B\rangle\,\frac{\langle 0_B|0_{BCS}\rangle}{\langle 0_B|0_B\rangle}\hspace{2cm}\nonumber\\
+P_{\perp}\,\frac{1}{\hat{\mathcal{E}}_{BCS}-\hat{H}_B}\,P_{\perp}\, W|0_{BCS}\rangle\
\end{eqnarray}
its iteration leading to
\begin{equation}
|0_{BCS}\rangle=(1+Q_{\perp}W)|0_B\rangle\frac{\langle 0_B|0_{BCS}\rangle}{\langle 0_B|0_B\rangle}\ ,
\end{equation}
where the operator $Q_{\perp}$, defined as
\begin{eqnarray}
Q_{\perp}=\sum_{n=0}^{+\infty}\left(P_{\perp}\frac{1}{\hat{\mathcal{E}}_{BCS}-\hat{H}_B}P_{\perp}W\right)^n\nonumber\\
\times\ P_{\perp}\frac{1}{\hat{\mathcal{E}}_{BCS}-\hat{H}_B}P_{\perp}\ ,
\end{eqnarray}
acts in the subspace perpendicular to the $\hat{H}_B$  ground state only. So, the denominators in the above sum stay finite, even for $\hat{\mathcal{E}}_{BCS}$ replaced by $\hat{\mathcal{E}}_B$.

The $\hat{H}_{BCS}$ ground state in Eq.(60) depends on $W$ in an explicit way but also in a hidden way through the ground state energy $\hat{\mathcal{E}}_{BCS}$. To get a similar expansion for
$\hat{\mathcal{E}}_{BCS}$, we project Eq.(54) onto $\langle 0_B|$. This gives
\begin{equation}
\hat{\mathcal{E}}_{BCS}=\hat{\mathcal{E}}_B+\frac{\langle 0_B|W|0_{BCS}\rangle}{\langle 0_B|0_{BCS}\rangle}\ .
\end{equation}
We then replace $|0_{BCS}\rangle$ in $\langle 0_B|W|0_{BCS}\rangle$ by Eq.(61). This ultimately gives the equation fulfilled  by $\hat{\mathcal{E}}_{BCS}$ as
\begin{eqnarray}
\hat{\mathcal{E}}_{BCS}=\hat{\mathcal{E}}_B+\frac{\langle 0_B|W|0_B\rangle}{\langle 0_B|0_B\rangle}\hspace{1.5cm}\nonumber\\
+\frac{\langle 0_B|W\,Q_{\perp}\,W|0_{B}\rangle}{\langle 0_B|0_B\rangle}\ .
\end{eqnarray}
Note that $\hat{\mathcal{E}}_{BCS}$ also is in the RHS of the above equation through the $(\hat{\mathcal{E}}_{BCS}-\hat{H}_B)$ factors contained in $Q_{\perp}$. The above equation still shows that $\hat{\mathcal{E}}_{BCS}-\hat{\mathcal{E}}_B$ is first order in $W$; so, up to the second order in $W$, Eqs.(62) and (64) give
\begin{eqnarray}
\hat{\mathcal{E}}_{BCS}=\hat{\mathcal{E}}_B+\frac{\langle 0_B|W|0_B\rangle}{\langle 0_B|0_B\rangle}\hspace{1.5cm}\nonumber\\
+\frac{1}{\langle 0_B|0_B\rangle}\langle 0_B|WP_{\perp}\frac{1}{\hat{\mathcal{E}}_B-\hat{H}_B}P_{\perp}W|0_B\rangle\nonumber\\
+0(W^3)\ .
\end{eqnarray}

As obvious from these equations, the calculation of $\hat{\mathcal{E}}_{BCS}$ and $|0_{BCS}\rangle$ requires the knowledge of $W|0_B\rangle$. Since the $\hat{H}_B$ ground state $|0_B\rangle$ is such that $\tilde{a}_{\v k}|0_B\rangle=0=\tilde{b}_{-\v k}|0_B\rangle$ while the $W$ potential given in Eq.(18) reads in terms of $(B_{\v k}-z_{\v k})$, it is easy to calculate $W|0_B\rangle$ by first rewriting $B_{\v k}=b_{-\v k}a_{\v k}$ in terms of the Bogoliubov operators $\tilde{a}_{\v k}^\dag$ and $\tilde{b}_{-\v k}^\dag$. From Eqs.(24), we get
\begin{eqnarray}
a_{\v k}^\dag&=&\left(x_{\v k}'^\ast \tilde{a}_{\v k}^\dag-y_{\v k}\tilde{b}_{-\v k}\right)/D_{\v k}\ ,\nonumber\\
b_{-\v k}&=&\left(x_{\v k} \tilde{b}_{-\v k}-y_{\v k}'^\ast \tilde{a}_{\v k}^\dag\right)/D_{\v k}\ ,
\end{eqnarray}
with $D_{\v k}=x_{\v k}x_{\v k}'^\ast-y_{\v k}y_{\v k}'^\ast=x_{\v k}x_{\v k}'^\ast\left(1+|f_{\v k}|^2\right)$, due to Eq.(27); so, $|D_{\v k}|=1$, which follows from Eq.(28). As a result, Eqs.(27,28) lead to
\begin{equation}
B_{\v k}-z_{\v k}=Z_{\v k}+T_{\v k}\tilde{N}_{\v k}+X_{\v k}\tilde{B}_{\v k}+Y_{\v k}^\ast\tilde{B}_{\v k}^\dag\ .
\end{equation}
$\tilde{B}_{\v k}^\dag=\tilde{a}_{\v k}^\dag \tilde{b}_{-\v k}^\dag$
creates a pair of Bogoliubov excitations, while $\tilde{N}_{\v
k}=\tilde{a}_{\v k}^\dag \tilde{a}_{\v k}+\tilde{b}_{-\v k}^\dag
\tilde{b}_{-\v k}$ is the number operator for these excitations. The
prefactors in Eq.(67) are given by $Z_{\v k}=-x_{\v k}y_{\v
k}^\ast-z_{\v k}$, $T_{\v k}=x_{\v k}y_{\v k}^\ast$, $X_{\v k}=x_{\v
k}x'_{\v k}$ and $Y_{\v k}=y_{\v k}y'_{\v k}$.

We are going to show that the appropriate choice for the $z_{\v k}$ scalars introduced in the Bogoliubov procedure corresponds to set $Z_{\v k}=0$.

\subsection{First-order correction to the Bogoliubov energy}

The first-order correction to the Bogoliubov energy $\hat{\mathcal{E}}_B$ is given by the second term of Eq.(65). It precisely reads, using Eq.(18),
\begin{eqnarray}
\frac{\langle 0_B|W|0_B\rangle}{\langle 0_B|0_B\rangle}=\hspace{4.5cm}\nonumber\\
-\sum_{\v k'}\sum_{\v k}V_{\v k'\v k}\frac{\langle 0_B|(B_{\v k'}^\dag-z_{\v k'}^\ast)(B_{\v k}-z_{\v k})|0_B\rangle}
{\langle 0_B|0_B\rangle}\ .
\end{eqnarray}
Since $\tilde{N}_{\v k}|0_B\rangle$ and $\tilde{B}_{\v k}|0_B\rangle$ are both equal to zero, Eq.(67) leads to $(B_{\v k}-z_{\v k})|0_B\rangle=Z_{\v k}|0_B\rangle+Y_{\v k}^\ast \tilde{B}_{\v k}^\dag|0_B\rangle$. So, the first-order correction to the Bogoliubov energy $\hat{\mathcal{E}}_B$ reduces to
\begin{eqnarray}
\frac{\langle 0_B|W|0_B\rangle}{\langle 0_B|0_B\rangle}=\hspace{4,5cm}\nonumber\\
-\sum_{\v k'}\sum_{\v k}V_{\v k'\v k}Z_{\v k'}^\ast Z_{\v k}
-\sum_{\v k}V_{\v k\v k}|Y_{\v k}|^2\ .
\end{eqnarray}
The second term in the RHS of the above equation is sample volume free because the potential $V_{\v k'\v k}$ depends on sample volume as $1/L^3$ while the $\v k$ sum brings a $L^3$ factor. By contrast, the first term of Eq.(69) has a double $\v k$ sum which brings a $L^6$ factor; so, it linearly increases with sample volume. This first term would thus give an extensive difference between the BCS energy $\hat{\mathcal{E}}_{BCS}$ and its Bogoliubov value $\hat{\mathcal{E}}_B$. However, it is possible to cancel this difference by setting $Z_{\v k}=0$, i.e., by choosing the $z_{\v k}$ scalars as
\begin{equation}
z_{\v k}=-x_{\v k}y_{\v k}^\ast=-\frac{f_{\v k}^\ast}{1+|f_{\v k}|^2}=\frac{\Delta_{\v k}}{2\tilde{E}_{\v k}}\ ,
\end{equation}
according to Eqs.(69) and (34). When inserted into Eq.(20), this gives
\begin{equation}
\Delta_{\v k'}=\sum_{\v k}V_{\v k'\v k}\,z_{\v k}=\sum_{\v k}V_{\v k'\v k}\,\frac{\Delta_{\v k}}{2\tilde{E}_{\v k}}\ ,
\end{equation}
with $\tilde{E}_{\v k}$ defined in Eq.(36).

To go further and get some explicit results, we must choose a particular form for the $V_{\v k'\v k}$ potential. By taking this potential as constant and separable, namely,
\begin{equation}
V_{\v k'\v k}=V\omega_{\v k'}\omega_{\v k}\ ,
\end{equation}
with $\omega_{\v k}^2=\omega_{\v k}$, the solution of Eq.(71) readily gives
\begin{eqnarray}
\Delta_{\v k'}=V\omega_{\v k'}\sum_{\v k}\omega_{\v k}\frac{\Delta_{\v k}}{2\tilde{E}_{\v k}}\
=\omega_{\v k'}\,\Delta.
\end{eqnarray}
If we now combine Eqs. (71) and (73), we get $\Delta$ through
\begin{equation}
\frac{2}{V}=\sum_{\v k}\frac{\omega_{\v k}}{\tilde{E}_{\v k}}=\sum_{\v k}\frac{\omega_{\v k}}{\sqrt{\xi_{\v k}^2+|\Delta|^2}}\ ,
\end{equation}
which is known as the ``gap equation''. Note that, in order for the RHS of the above equation to be positive,  $\tilde{E}_{\v k}$ must be positive, in agreement with Eq.(36).

In the following, we are going to note as $W_B$ the two-body interaction $W$ in which the $z_{\v k}$ scalars are given by Eq.(70). This two-body interaction splits in terms of Bogoliubov operators as
\begin{eqnarray}
W_B=-VU_B^\dag U_B\hspace{2cm} \nonumber\\
U_B=\sum_{\v k}\left(T_{\v k}\tilde{N}_{\v k}+X_{\v k}\tilde{B}_{\v k}+Y_{\v k}^\ast\tilde{B}_{\v k}^\dag\right)\ .
\end{eqnarray}
Since $U_B|0_B\rangle=\sum_{\v k}Y_{\v k}^\ast\tilde{B}_{\v
k}^\dag|0_B\rangle$, we easily find that $P_{\perp}W_B|0_B\rangle$ reduces to states having one or two pairs of Bogoliubov
excitations
\begin{eqnarray}
P_{\perp}W_B|0_B\rangle=-V\sum_{\v k_1}P_{\v k_1}\tilde{B}_{\v k_1}^\dag|0_B\rangle\hspace{1.5cm}\nonumber\\
-V\sum_{\v k_1\neq\v k_2}P_{\v k_1\v k_2}\tilde{B}_{\v k_1}^\dag\tilde{B}_{\v k_2}^\dag|0_B\rangle\ ,
\end{eqnarray}
with $P_{\v k_1}=2T_{\v k_1}^\ast Y_{\v k_1}^\ast$ and $P_{\v k_1\v k_2}=X_{\v k_1}^\ast Y_{\v k_2}^\ast$.

\subsection{Higher-order corrections}

Let us now consider the second order correction to the $\hat{\mathcal{E}}_{BCS}$ energy, given by the third term of Eq.(65). Using Eqs (31) (48) and (76), we find
\begin{eqnarray}
\frac{\langle 0_B|W_BP_{\perp}\left(\hat{\mathcal{E}}_B-\hat{H}_B\right)^{-1}P_{\perp}W_B|0_B\rangle}{\langle 0_B|0_B\rangle}=\hspace{1cm}\nonumber\\
V^2\sum_{\v k_1}\frac{|P_{\v k_1}|^2}{-2\tilde{E}_{\v k_1}}+V^2\sum_{\v k_1\neq\v k_2}
\frac{|P_{\v k_1\v k_2}|^2+|P_{\v k_2\v k_1}|^2}{-2\tilde{E}_{\v k_1}-2\tilde{E}_{\v k_2}}.
\end{eqnarray}

The first term of this second-order correction contains two $V$'s and one $\mathbf{k}$ sum only; so, it
goes to zero with sample volume as $1/L^3$. The second term, which also has two $V$'s but
two $\mathbf{k}$ sums, is sample volume free. However,
it still gives correction to the $\hat{\mathcal{E}}_B$ ground
state energy smaller than $L^3$, i.e., underextensive and thus negligible in
the thermodynamical limit.

By counting the number of potentials $V$ and the number of $\mathbf{k}$ sums, it is possible
to show that all higher order terms of the $W_B$ expansion of the ground state
energy $\hat{\mathcal{E}}_{BCS}$ also give underextensive corrections. So, within the choice of $z_{%
\mathbf{k}}$ scalars given in Eq.(70), the extensive part of the ground state energy of the
BCS Hamiltonian $\hat{H}_{BCS}$ in the grand canonical ensemble, indeed reduces to  $%
\hat{\mathcal{E}}_B$ given in Eq.(38), i.e., to the ground state energy of the Bogoliubov Hamiltonian $\hat{H}_B$.

\section{Mean values of a few other physically relevant operators}

In the above section, we have shown that $\hat{\mathcal{E}}_{BCS}$
is equal to $\hat{\mathcal{E}}_B$ within underextensive
contributions, i.e.,
\begin{equation}
\frac{\langle 0_{BCS}|\hat{H}_{BCS}|0_{BCS}\rangle}{\langle 0_{BCS}|0_{BCS}\rangle}\simeq
\frac{\langle 0_B|\hat{H}_B|0_B\rangle}{\langle 0_B|0_B\rangle}\ .
\end{equation}
In this section, we are going to show that the mean values of a few physically relevant operators $A$ are the same when calculated in the $\hat{H}_{BCS}$ ground state or in the $\hat{H}_B$ ground state, within terms which are negligible in the large sample limit, i.e., $\langle A\rangle_{BCS}\simeq\langle A\rangle_B$ with
\begin{eqnarray}
\langle A\rangle_{BCS}&=&\frac{\langle 0_{BCS}|A|0_{BCS}\rangle}{\langle 0_{BCS}|0_{BCS}\rangle}\nonumber\\
\langle A\rangle_B&=&\frac{\langle 0_{B}|A|0_{B}\rangle}{\langle 0_{B}|0_{B}\rangle}\ .
\end{eqnarray}

Using Eq.(61) for $|0_{BCS}\rangle$ and Eq.(65) for $\hat{\mathcal{E}}_{BCS}$, one can show that the difference between these two mean values reads, up to second order in $W_B$, as
\begin{eqnarray}
\langle A\rangle_{BCS}-\langle A\rangle_B\simeq \frac{\langle 0_B|AQ_{\perp}W_B|0_B\rangle +\mathrm{c.c.}}{\langle 0_B|0_B\rangle}\hspace{0.5cm}\nonumber\\
+\frac{\langle 0_B|AQ_{\perp}W_BQ_{\perp}W_B|0_B\rangle+\mathrm{c.c.}}{\langle 0_B|0_B\rangle}
\nonumber\\
+\frac{\langle 0_B|W_BQ_{\perp}AQ_{\perp}W_B|0_B\rangle}{\langle 0_B|0_B\rangle}\hspace{1cm}\nonumber\\
-\frac{\langle 0_B|W_BQ_{\perp}^2W_B|0_B\rangle}{\langle
0_B|0_B\rangle}\langle A\rangle_B\ ,\hspace{1cm}
\end{eqnarray}
where $Q_{\perp}$, given in Eq.(62), is here reduced to its $n=0$ term.

\subsection{Electron distribution}

Let us consider the operator $A=a_{\v p}^\dag a_{\v p}$ which physically is a quite relevant operator because, from it,
we can probe the distribution of up or down spin electrons in the $|0_B\rangle$ and $|0_{BCS}\rangle$
ground states. Using Eq.(66), the number of up spin electrons with momentum $\v
p$ reads as a sum of three Bogoliubov operators:
\begin{eqnarray}
a_{\v p}^\dag a_{\v p}=|x'_{\v p}|^2\tilde{a}_{\v p}^\dag\tilde{a}_{\v p}+|y_{\v p}|^2\tilde{b}_{-\v p}\tilde{b}_{-\v p}^\dag\nonumber\\
 -\left(x'_{\v p}y_{\v p}\tilde{B}_{\v p}+\mathrm{h.c.}\right).
\end{eqnarray}

We are going to show that difference in the mean values of each of these Bogoliubov operators calculated in $|0_{BCS}\rangle$ and $|0_B\rangle$ vanishes in the large sample limit. Let us start with $\tilde{A}=\tilde{a}_{\v p}^\dag\tilde{a}_{\v p}$. Since $\tilde{A}|0_B\rangle=0$, we readily find $\langle\tilde{a}_{\v p}^\dag\tilde{a}_{\v p}\rangle_B=0$, while the four terms of Eq.(80), with Eq.(76) used for $P_{\perp}W_B|0_B\rangle$, reduce to
\begin{eqnarray}
\langle\tilde{a}_{\v p}^\dag\tilde{a}_{\v p}\rangle_{BCS}\simeq\frac{\langle 0_B|W_BQ_{\perp}\tilde{A}Q_{\perp}W_B|0_B\rangle}
{\langle 0_B|0_B\rangle}\hspace{0.5cm}\nonumber\\
\simeq V^2\frac{|P_{\v p}|^2}{4\tilde{E}_{\v p}^{2}}+V^2\sum_{\v
k\neq\v p}\frac{|P_{\v p\v k}|^2+|P_{\v k\v p}|^2}{2(\tilde{E}_{\v
p}+\tilde{E}_{\v k})^{2}},
\end{eqnarray}
 The first term of this mean value goes to zero with sample volume as $1/L^6$ while the second term goes to zero as $1/L^3$. Thus, these two terms give a negligible contribution in the large sample limit.

 To get the mean value of $\tilde{b}_{-\v p}\tilde{b}_{-\v p}^\dag$, we first rewrite this operator as $1-\tilde{b}_{-\v p}^\dag\tilde{b}_{-\v p}$ and then use the result for $\tilde{a}_{\v p}^\dag\tilde{a}_{\v p}$.

 We now turn to the last term of Eq.(81), namely $\tilde{R}_{\v p}=t\tilde{B}_{\v p}+\mathrm{h.c.}$, with $t=-x'_{\v p}y_{\v p}$. Using Eqs.(76) and (80), the first order term in $W_B$ follows from
\begin{equation}
\langle 0_B|\tilde{R}_{\v p}P_{\perp}\frac{1}{\hat{\mathcal{E}}_B-\hat{H}_B}P_{\perp}W_B|0_B\rangle=-Vt\frac{P_{\v p}}{-2\tilde{E}_{\v p}},
\end{equation}
which goes to zero as $V \sim 1/L^3$. If we now
consider the $W_B^2$ terms, we find that they contain
$V^2$ and a single sum over $\v k\neq\v p$ at most; so, these second
order terms also go to zero with sample volume as $1/L^3$. We thus
end with
\begin{equation}
\langle a_{\v p}^\dag a_{\v p}\rangle_{BCS}\simeq \langle a_{\v p}^\dag a_{\v p}\rangle_{B}=|y_{\v p}|^2=\frac{|f_{\v p}|^2}{1+|f_{\v p}|^2},
\end{equation}
within terms in $1/L^3$. This shows that, in the thermodynamical limit, the electron distribution in the BCS
ground state $|0_{BCS}\rangle$ is just the same as in the BCS ansatz since, as shown above, this ansatz is identical to the Bogoliubov Hamiltonian ground state $|0_B\rangle$.

\subsection{Other physical quantities}

Similar calculations can be done for other physical quantities. To perform these calculations in an easy way, we first transform the operators which represent these physical quantities, commonly written in terms of $a_{\v p}^\dag$ and $b_{-\v p'}^\dag$, into hermitian operators written in terms of $\tilde{a}_{\v p}^\dag$ and $\tilde{b}_{-\v p'}^\dag$. We then separately show that the mean value difference of operators like $(\tilde{a}_{\v p_1}^\dag\tilde{a}_{\v p_2}^\dag \tilde{a}_{\v p_2}\tilde{a}_{\v p_1})$ or $(\tilde{B}_{\v p_1}^\dag\tilde{B}_{\v p_2}^\dag+\mathrm{h.c.})$ goes to zero in the large sample limit, by counting the number of $V$ factors and the number of free $\v k$ sums, as we have done for $\tilde{a}_{\v p_1}^\dag\tilde{a}_{\v p_1}$ or $(\tilde{B}_{\v p_1}^\dag+\tilde{B}_{\v p_1})$: we can show that we always have one more $V$ than the number of free $\v k$ sums, so that the associated terms indeed are underextensive.

For all operators we have considered, we ended with
\begin{equation}
\langle A\rangle_{BCS}=\langle A\rangle_B\ [1+0(1/L^3)]\ .
\end{equation}
This result is definitely astonishing because, in the BCS ansatz, all pairs are condensed into
the same state; so, this ansatz is fundamentally different from the exact Richardson-Gaudin form of the BCS ground state. The result of Eq.(85) seems to indicate that this formal difference is physically irrelevant in the large sample limit since it gives negligible contributions to the mean values of physical quantities calculated either within the $\hat{H}_{BCS}$ ground state or within the $\hat{H}_B$ ground state---which just corresponds to the BCS ansatz.

\section{Extensive part of the $H_{BCS}$ ground state energy}

We have shown in section V that the extensive parts of the $\hat{H}_{BCS}$ and $\hat{H}_{B}$ ground state energies are equal, $\hat{\mathcal{E}}_{BCS}\simeq\hat{\mathcal{E}}_B$, when the $z_{\v k}$ scalars are chosen according to Eq.(70). The ground state energy $E_N$ of the
$H_{BCS}$ Hamiltonian in the \emph{canonical} ensemble is related to the $\hat{H}_{BCS}$ energy $\hat{\mathcal{E}}_{BCS}$ through $\hat{\mathcal{E}}_{BCS}=E_N-2\mu N$, where $N$ is equal to the mean value $\overline{N}$ of the fermion-pair number in the system at hand. For a potential taken in the separable form of Eq.(72) in order to possibly reach analytical results, we then find, using Eqs.(20,72,73)
\begin{equation}
\sum_{\mathbf{k}^{\prime},\mathbf{k}}V_{\mathbf{k}^{\prime}\mathbf{k}}z_{%
\mathbf{k}^{\prime}}^\ast z_{\mathbf{k}}=V\sum_{\mathbf{k}^{\prime}}z_{%
\mathbf{k}^{\prime}}^\ast\omega_{\mathbf{k}^{\prime}}\sum_{\mathbf{k}}z_{%
\mathbf{k}}\omega_{\mathbf{k}}=\frac{|\Delta|^2}{V}\ .
\end{equation}
This gives the BCS ground state energy $E_N$ as
\begin{equation}
E_N=\hat{\mathcal{E}}_B+2\mu \overline{N}=\frac{|\Delta|^2}{V}+\sum_{\mathbf{k}}(\xi_{\mathbf{k}%
}-\tilde{E}_{\mathbf{k}})\omega_{\mathbf{k}}+2\mu \overline{N}\ .
\end{equation}
Note that an additional factor $\omega_{\mathbf{k}}$ can freely be introduced
in the $\mathbf{k}$ sum of the above equation because  $\xi_{\mathbf{k} }-\tilde{E}_{\mathbf{k}}$ is equal to
0 outside the potential layer due to Eq.(73). The reason for
introducing this $\omega_{\mathbf{k}}$ factor is to possibly split the
$\mathbf{k}$ sum and calculate its two parts separately within the
potential layer.

The chemical potential $\mu$, hidden in $\xi_{%
\mathbf{k}}$, follows from enforcing  the pair number mean value to be equal to the number of pairs $N$ at hand
\begin{eqnarray}
N=\overline{N}=\frac{\langle 0_B|\sum_{\v k}a_{\v k}^\dag a_{\v k}|0_B\rangle}{\langle 0_B|0_B\rangle}=
\sum_{\v k}\frac{|f_{\v k}|^2}{1+|f_{\v k}|^2}\nonumber\\
=\sum_{\mathbf{k}}\frac{\omega_{\mathbf{k}}}{2}\left(1-\frac{\xi_{\mathbf{k}%
}}{\tilde{E}_{\mathbf{k}}}\right)\ ,
\end{eqnarray}
due to Eqs.(84) and (35).

\subsection{Standard BCS configuration}

In the standard BCS configuration, the potential layer extends between $%
\epsilon_{F_0}$ and  $\epsilon_{F_0}+\Omega$ in a region where the
density of states can be taken as equal to a constant $\rho$. Moreover, this potential is said to extend symmetrically with respect to the normal electron Fermi sea $\epsilon_F$. This corresponds to have electron pairs filling half of the potential layer, i.e., $N$ equal to $\rho\Omega/2=N_{BCS}$.

For half-filling and a constant density of states $\rho$, Eq.(88)
gives, by turning to the continuous limit as valid in the large sample limit since $1/\rho$ is as small as $1/L^3$
\begin{equation}
\frac{\rho\Omega}{2}=\int_{\epsilon_{F_0}-\mu}^{\epsilon%
_{F_0}-\mu+\Omega}\rho\, d\xi\,\frac{1}{2} \left(1-\frac{\xi}{\sqrt{%
\xi^2+\Delta^2}}\right)\ .
\end{equation}
It is then straightforward to check that this equation is fulfilled for $\mu=\epsilon_{F_0}+\Omega/2$.

The gap equation (74), for a chemical
potential set in the middle of the potential layer,  then reads
\begin{equation}
\frac{2}{V}=\int_{-\Omega/2}^{\Omega/2}\frac{\rho\,d\xi}{\sqrt{\xi^2+\Delta^2
}}=\rho\log\frac{\frac{\Omega}{2}+\sqrt{\frac{\Omega^2}{4}+\Delta^2}}{-\frac{
\Omega}{2}+\sqrt{\frac{\Omega^2}{4}+\Delta^2}}.
\end{equation}
This gives
\begin{equation}
\Delta=\frac{\Omega\,e^{-1/\rho V}}{1-e^{-2/\rho V}}=\frac{\Omega\sqrt{\sigma
}}{1-\sigma}\ ,
\end{equation}
where we have set $\sigma=e^{-2/\rho V}$.

For $\mu=\epsilon_{F_0}+\Omega/2$, the two sums over $\v k$ in Eq.(87) reduce respectively to $\sum_{
\mathbf{k}}\xi_{\mathbf{k}}\omega_{\mathbf{k}}=0$ and to
\begin{eqnarray}
\sum_{\mathbf{k}}\tilde{E}_{\mathbf{k}}\omega_{\mathbf{k}}
&=&\int_{-\Omega/2}^{\Omega/2}\rho\,d\xi\sqrt{\xi^2+\Delta^2}\nonumber\\
&=&
\frac{\rho\Omega}{2}\sqrt{\frac{\Omega^2}{4}+\Delta^2}+\frac{\Delta^2}{V}\ ,
\end{eqnarray}
as obtained from an integration by parts.
If we now insert this result into Eq.(87),  we end with a ground state energy in the canonical ensemble given by
\begin{equation}
E_{N_{BCS}}=2N\epsilon_{F_0}+N\Omega-\frac{\rho\Omega}{2}\sqrt{%
\frac{\Omega^2}{4}+\Delta^2}\ .
\end{equation}

Since $\Delta=0$ for $V=0$, the ground state energy in the absence of
potential $E^{(0)}_{N_{BCS}}$ reduces to $2N\epsilon%
_{F_0}+N\Omega-\rho\,\Omega^2/4$. So, the condensation energy induced by
the attracting potential $V$ is given by
\begin{equation}
E^{(0)}_{N_{BCS}}-E_{N_{BCS}}=\frac{\rho\,\Omega}{2}\left(%
\sqrt{\frac{\Omega^2}{4}+\Delta^2}-\frac{\Omega}{2}\right)\ .
\end{equation}
For $\Delta$ given by Eq.(91), we find that $\sqrt{1+4\Delta^2/\Omega^2}$ reduces
to $(1+\sigma)/(1-\sigma)$. This shows that the condensation energy in the case of half filling ends by taking a quite compact form
\begin{equation}
E^{(0)}_{N_{BCS}}-E_{N_{BCS}}=\frac{\rho\,\Omega}{2}\,%
\frac{\Omega\sigma}{1-\sigma}=\frac{1}{2} \rho\,\Delta^2(1-\sigma)\ .
\end{equation}

It is of interest to note that this result fully agrees with the expression of the $N$-pair energy we have
obtained by analytically solving Richardson-Gaudin equations in the
canonical ensemble \cite{Crouzeix}, namely,
\begin{equation}
E_N=NE_1+\frac{N(N-1)}{\rho}\,\frac{1+\sigma}{1-\sigma}\ ,
\end{equation}
where $E_1=2\epsilon%
_{F_0}-2\Omega\sigma/(1-\sigma)$ is the single pair energy obtained by Cooper. Indeed, the above equation gives the
condensation energy, i.e., the part of $E_N$ which cancels when $\sigma=0$, as
\begin{equation}
E^{(0)}_N-E_N=N\left(1-\frac{N}{\rho\Omega}\right)\frac{2\sigma}{1-\sigma}%
\,\Omega\ .
\end{equation}
It is then easy to check that this condensation energy reduces to Eq.(95) for $N=N_{BCS}=\rho\Omega/2$, whatever $\sigma$, i.e., not only in the weak coupling limit, $\sigma\simeq0$.

\subsection{Arbitrary filling}

We now consider an arbitrary filling of the potential layer, namely,
$N$ not exactly equal to half the number of states between
$\epsilon_{F_{0}}$ and $\epsilon_{F_{0}}+\Omega $. This situation
can be considered as a toy model for the density-induced crossover
between the regimes of isolated fermionic molecules and dense BCS
condensate. Such a model can also be relevant to situations
encountered in semiconductors \cite{Geyer,Eagles}, where the density
of carriers forming correlated pairs can be low. Moreover, it can be
tuned by changing the doping. As a result, at very low density of
correlated pairs, the chemical potential can go below the bottom of
the band which, in this paper, is referred as the potential layer.
This scenario is similar to the situation, considered below.

In the case of an arbitrary filling of the potential layer, the
chemical potential $\mu$ does not necessarily fall exactly in the
middle of the potential layer, as it happens in the
standard BCS configuration considered above. This means that we now
have to determine the gap $\Delta$ and the chemical potential $\mu$
simultaneously by solving two coupled equations, namely, Eq.(74) for
$\Delta$ and Eq.(88) for $\mu$, since $\Delta$ and $\mu$ depend on
each other. After switching from sum to integral in these two
equations, the equation for the gap reads
\begin{equation}
\frac{2}{\rho V}=\ln \frac{\epsilon_{F_{0}}-\mu +\Omega +\sqrt{(\epsilon%
_{F_{0}}-\mu +\Omega )^{2}+\Delta ^{2}}}{\epsilon_{F_{0}}-\mu +\sqrt{(%
\epsilon_{F_{0}}-\mu )^{2}+\Delta ^{2}}}\ ,
\end{equation}%
while the equation for the chemical potential appears as
\begin{eqnarray}
\Omega -\frac{2N}{\rho }=\hspace{5cm}\nonumber\\
\sqrt{(\epsilon_{F_{0}}-\mu +\Omega )^{2}+\Delta ^{2}}-\sqrt{(\epsilon%
_{F_{0}}-\mu )^{2}+\Delta ^{2}}.
\end{eqnarray}%
The analytical solution for these two equations reads as%
\begin{equation}
\Delta =\sqrt{\frac{N}{\rho\Omega }\left(1 -\frac{N}{\rho\Omega }\right) }\frac{%
2\sqrt{\sigma }}{1-\sigma }\,\Omega\ ,
\end{equation}%
\begin{equation}
\mu-\epsilon_{F_0} =\frac{N}{\rho }-\left( 1 -\frac{2N}{\rho\Omega }%
\right) \frac{\sigma }{1-\sigma }\,\Omega\ .
\end{equation}%
as easy to check. Note that, as expected, this solution reduces to the values
of $\Delta $ and $\mu $ obtained for half-filling, i.e., for
$N=N_{BCS}=\rho \Omega /2$. In particular, we see that
$\mu $ then is exactly in the middle of the potential layer for any $V$.

By inserting these expressions of $\Delta$ and $\mu$ into the
BCS ground state energy $E_N$ given in Eq.(87), we find
\begin{equation}
E_N=2N\epsilon_{F_0}+\frac{N^2}{\rho }-2N\Omega\left(
1 -\frac{N}{\rho\Omega }\right) \frac{\sigma }{1-\sigma }\ .
\end{equation}%
Since the ground state energy in the absence of potential, i.e., for $\sigma=0$,
reduces to the first two terms of the above equation, we readily see that the condensation
energy for arbitrary filling, i.e., for $N$ not exactly equal to $\rho\Omega/2$, is given by%
\begin{eqnarray}
E^{(0)}_N-E_N&=&N\left(1-
\frac{N}{\rho\Omega }\right) \frac{2\sigma }{1-\sigma }\,\Omega  \nonumber \\
&=&\frac{\rho \Delta^2}{2}(1-\sigma)\ .
\end{eqnarray}
Again, this result fully agrees with the expression of the ground
state energy derived from Richardson-Gaudin equations for arbitrary
$N$\ and $\sigma $, as given in Eq.(97). It also is worth noting that
the expression of condensation energy in terms of $\Delta $,
as given in the second line in Eq.(103), is universal, i.e.,
independent of the filling of the potential layer.

The condensation energy,  given in Eq.(103), can be physically
understood as each of the $N$\ correlated pairs bringing its ``own
binding energy", this binding energy being linearly decreased
compared to one isolated Cooper pair by a factor $(1-N/\rho\Omega)$
due to the Pauli exclusion principle between the electrons from
which the correlated pairs are constructed. In particular, for the
standard BCS configuration, this ``pair binding energy" is equal to one half the binding energy of a single Cooper pair. This
provides another appealing understanding of the usual BCS result for the
condensation energy.

\section{Excited state energy}

Let us end this work by considering excited states in the case of
arbitrary filling. Their energies are formally given by
$E_N+n\tilde{E}_{\v k}$ with $n=1,2,\cdots$, and $\tilde{E}_{\v k}$ given by Eqs. (36) and (74). The energy of the lowest excited state
corresponds to $n=1$; so, the energy difference between this
state and the ground state is $\tilde{E}_{\v k}=\sqrt{\xi_{\v
k}^2+\Delta^2}$. The minimum of this
quantity, commonly called "gap", is reached for the lowest possible value of $\xi _{\mathbf{k}%
}^{2}=(\epsilon _{\mathbf{k}}-\mu )^{2}$.

The problem then is to
determine this lowest possible value. This can be done by noting
that the electron energy $\epsilon_{\v k}$ by construction falls
inside the energy layer $(\epsilon_{F_0},\epsilon_{F_0}+\Omega)$. By contrast, the chemical potential $\mu $ does not necessarily fall inside
this energy layer. For $\mu $ inside this layer, the lowest $\tilde{E}_{\v k}$ corresponds to $\epsilon
_{\mathbf{k}}=$ $\mu $: we then get the standard result
\begin{equation}
 \left( \tilde{E}_{k}\right) _{\min }=\Delta\ :
 \end{equation}
difference between the minimum excited state energy and the
ground state energy $E_N$ is equal to $\Delta$, as found in textbooks.
Note that in the half-filled
configuration, Eq. (101) gives $\mu$ equal to $\epsilon_{F_0}+\Omega /2$ for any
$\sigma $; so, $\epsilon_{\mathbf{k}}$ can be equal to $\mu$ and the
gap is equal to $\Delta$ for any $V$.

Let us consider the possibility of having $\mu$ outside of the energy
layer. Equation (101) shows that for $N$ smaller than
\begin{equation}
N_{1}=\rho \Omega \frac{\sigma }{1+\sigma }\ ,
\end{equation}%
the chemical potential $\mu$ falls below $\epsilon_{F_0}$,
while for $N$ larger than
\begin{equation}
N_{2}=\rho \Omega -\rho \Omega \frac{\sigma }{1+\sigma }=\rho \Omega -N_{1}\ ,
\end{equation}%
it falls above $\epsilon_{F_0}+\Omega$; so,
for $N<N_1$ or $N_2<N$, we cannot have $\epsilon _{\mathbf{k}}=$ $%
\mu $.

Since the number of pairs inside the
potential layer is $\rho\Omega$ , the above equations
show that the intervals of $N$ for these two ``anomalous" phases,
$0<N<N_1$ and $N_2<N<\rho\Omega$, are exactly as large, but still
very narrow in the weak-coupling limit, $\sigma\ll 1$.

For $0<N<N_{1}$, the minimum value of
$(\epsilon _{\mathbf{k}}-\mu )^{2}$ is reached for the lowest possible
$\epsilon _{\mathbf{k}}$, namely $\epsilon
_{\mathbf{k}}=\epsilon_{F_{0}}$. Using Eqs.(35,100,101), the energy
gap associated to the $\tilde{E}_{k}$ minimum is then given, for $N$ between 0 and $N_1$, by
\begin{equation}
\Delta^{(dilute)}=\frac{N}{\rho }+\Omega \frac{\sigma }{1-\sigma }
\end{equation}%
The first term in the RHS of the above equation does not depend on
$V$. We can understand this $\sigma=0$ term by noting that, for $N$ between 0 and $N_1$, the
excitation energy minimum corresponds to
$\epsilon_{\v k}=\epsilon_{F_0}$.
 We then create a hole at $\epsilon_{\v k}=\epsilon_{F_0}$ and, in the absence of interaction, eject the
corresponding electron to the top of the non-interacting electron
Fermi sea, which for $N$ electrons in the
potential layer, corresponds to an energy equal to
$\epsilon_{F_0}+N/\rho$ when the density of states is constant and
equal to $\rho$. The energy difference between this excited state
and the ground state then is $N/\rho$, in agreement with Eq.(107)
taken for $\sigma= 0$.
The second term of  $\Delta^{(dilute)}$ is just half the binding energy of a
single pair, as obtained by Cooper.

Similarly, for $N_2<N<\rho\Omega$, the minimum of the excitation energy is
reached for $\epsilon_{\v k}=\epsilon_{F_0}+\Omega$. This minimum energy corresponds
to a gap given by
\begin{equation}
\Delta^{(dense)}=\left(\Omega -\frac{N}{\rho }\right)+\Omega \frac{\sigma }{%
1-\sigma }.
\end{equation}%
The first term of $\Delta^{(dense)}$ does not depend on $V$. It corresponds to an
electron ejected from the top of the normal Fermi sea which for $N$ electron pairs corresponds to $\epsilon_{F_0}+N/\rho$,  to the top of
the potential layer, i.e., $\epsilon_{F_0}+\Omega$. As a result, the energy difference for $\sigma= 0$ must be equal
to $(\epsilon_{F_0}+\Omega)-(\epsilon_{F_0}+N/\rho)$, i.e.,
$\Omega-N/\rho$, in agreement with the first term of Eq.(108). The
second term of this equation again is equal to half the binding
energy of the pair, as obtained in the dilute case, Eq.(107).

The gaps in the super dilute and super dense limits, given in Eqs.(107) and (108),
nicely show the duality which exists between
electrons and holes in the potential layer: for $%
0<N<N_{1}$, bound pairs are formed out of electrons, while for
$N_2<N<\rho\Omega$, one can think on pairs formed out of holes,
i.e., empty states in the layer. In order to excite the system, one
has to break a pair; this explains why, in the minimum of $\tilde{E}_{\v
k}$, appears one half of the single pair binding energy, a full
``pair breaking'' corresponding to \textit{two} excitations, not just
one.

The existence of a duality between electrons and holes also shows up
for $N_1<N<N_2$, when the gap is $\left(
\tilde{E}_{k}\right)_{\min}=\Delta$: as seen from Eq.(104),
the situation then is fully symmetrical with respect to the mutual
exchange of electron and hole numbers.

(i) For $\rho V\ll 1$, i.e., in the weak-coupling limit,
$N_{1}$ and $\rho \Omega -N_{2}$ correspond to the number
of states within half the single pair binding energy.
$N<N_{1}$ can be seen as a ``superdilute'' regime of pairs or a
``superdense" regime of holes, the wave functions of individual
pairs overlapping only slightly. By contrast, $N_2<N$ corresponds to a
``superdilute regime of holes'', or a ``superdense" regime of
electrons.

(ii) When $\rho V$ increases, $N_{1}$ approaches $\rho \Omega /2$
from below, while $N_{2}$ approaches $\rho \Omega /2$ from above: the interval of $N$'s which corresponds to the
usual gap $ \left( \tilde{E}_{k}\right) _{\min } =
\Delta $ shrinks when $\rho V$ increases, down to
zero when $\rho V$ infinite.

 We wish
to note that transitions at $N=N_{1}$ or $N=N_{2}$ should be
smooth with respect to the \emph{ground state} energy because the
same expression (102) holds in the three regimes, $N<N_1$,
$N_1<N<N_2$ and $N_2<N$. By contrast, in the case of the \emph{first
excited states}, these three regimes correspond to
\textit{different} kinetic energies for the excitation, as seen from
Eqs.(100,107,108). This difference induces discontinuities in
higher-order derivatives of $\left( \tilde{E}_{k}\right) _{\min }$
with respect to $\rho V$, on both sides of $N_1$ and $N_2$.

The configuration considered here, with a pair number $N$ being a
free parameter not exactly equal to $N_{BCS}$, i.e., to
half-filling, can be seen as a \textit{density-induced}
crossover model between isolated fermionic molecules for $N$ very
small, towards a dense regime of Cooper pairs, in which BCS
superconductivity exists. In the dilute regime of pairs, the
excitation energy is controlled by the single pair binding energy,
as we find, while at higher densities, it is controlled by a
cooperative many-body effect. The analysis
presented in this section has some similarities with the BEC-BCS
crossover considered by Eagles\cite{Eagles} and also by
Leggett\cite{Tonycross}. As main differences, Eagles keeps a non-constant density of states in
$\sqrt{\epsilon_{\v k}}$, while Leggett does not use an upper sharp
cut-off at $\epsilon_{F_{0}}+\Omega $ for the potential, irrelevant divergences being cured by a
``regularization" procedure.

The interesting aspect of the
model we here consider, with a upper cut-off and a constant
density of states ---   2D superconductors having constant density of states --- is that it
leads to simple analytical results. The upper cut-off at
$\epsilon_{F_{0}}+\Omega $ nicely evidences the existence of a ``superdense"
regime of pairs for $N$ close to complete filling $\rho \Omega $,
this regime being understood as a dilute regime of holes in the
potential layer with similarities with the dilute regime of
electrons when $N$ is very small.

\section{Concluding comments}

In this paper, we show that, through a proper choice of the $z_{\v
k}$ scalars introduced in the Bogoliubov procedure, it is possible
to reduce the difference between the ground state energies of the
BCS Hamiltonian $\hat{H}_{BCS}$ and the Bogoliubov Hamiltonian
$\hat{H}_B$ to underextensive terms, negligible in the
thermodynamical limit. The energy of the $\hat{H}_B$ ground state
fully agrees with the value obtained by analytically solving
Richardon-Gaudin equations which give the \emph{exact} eigenstates
of the BCS Hamiltonian in the canonical ensemble. Actually, this
agreement stays quite puzzling because the ground state of the
Bogoliubov Hamiltonian corresponds to the wave function ansatz
proposed by Bardeen, Cooper and Schrieffer, \emph{with all pairs
condensed into the same state}, while the exact Richardon-Gaudin
eigenstate reads as a product of correlated pair states which are
\emph{by construction all different due to the Pauli exclusion
principle}.

To better understand this point, we have considered mean
values of a few relevant operators, in particular the distribution
of electrons having a momentum $\v p$. Through an expansion of the
$\hat{H}_{BCS}$ ground state in $\hat{H}_{BCS}-\hat{H}_B$, we can
show that the $z_{\v k}$ scalars which lead to underextense
corrections to the ground state energy at any order in
$\hat{H}_{BCS}-\hat{H}_B$, lead to a similar result for the electron
number mean value. This indicates that, although the BCS ansatz is
formally different from the exact BCS ground state, this
difference does not seem to show up through physically relevant
quantities in the large sample limit.

In this work, we have also considered an arbitrary number of
pairs in the BCS potential. We show that the BCS ansatz not only gives
the exact energy for arbitrary filling, as obtained from analytically
solving Richardson-Gaudin equations, but it also allows us to evidence the
existence of a super dilute and a super dense regime of pairs in which
\emph{the energy gap is different from its usual value} $\Delta $. An electron/hole symmetry
is shown to exist between these two interesting regimes.

So, at the present time, we have two formally different forms for
the ground state of $N$ up and down spin electron pairs, which exactly give the same
extensive part of the ground state energy:

 (i) the BCS ansatz which
also is the Bogoliubov Hamiltonian ground state, its projection on
the $N$-pair subspace reading as
\begin{equation}
|\psi_{BCS}\rangle=B^{\dag N}|0\rangle\ ,
\end{equation}
where $B^\dag=\sum_{\v k}B_{\v k}^\dag (v_{\v k}/u_{\v k})$ with
$B_{\v k}^\dag =a_{\v k}^\dag b_{-\v k}^\dag$. Note that this gives
a pair wave function equal to $v_{\v k}/u_{\v k}$ which definitely
differs from its commonly quoted value $v_{\v k}u_{\v k}^\ast$ (for
more details, see Ref. \cite{Mon}).

(ii) the exact Richardson-Gaudin ground state which reads as
\begin{equation}
|\psi(N)\rangle=B^\dag(R_1)B^\dag(R_2)\cdots B^\dag(R_N)|0\rangle\ ,
\end{equation}
where $B^\dag(R_i)=\sum_{\v k}B_{\v k}^\dag [\omega_{\v
k}/(2\epsilon_{\v k}-R_i)]$, the $R_i$'s being by construction all
different due to Pauli blocking.

It is clear that, when $N$
increases, each of these states get more and more different from a
bare juxtaposition of either $B^\dag$ or $B^\dag(R)$ correlated
states. As a result, one cannot at the present time totaly exclude that, when $N$ gets
very large as in the thermodynamical limit, these two states end by
corresponding to the same linear combination of $B^\dag_{\v
k_1}B^\dag_{\v k_2}\cdots B^\dag_{\v k_N}$ products. Because the precise
microscopic understanding of BCS superconductivity goes along the
knowledge of the correct ground state, a precise study of the link between the
BCS ansatz and the exact Richardson-Gaudin ground state seems to
us as highly desirable. We hope that the present work will contribute to help
reopening a field commonly considered as fully understood.

\acknowledgments

This work is supported by the CNRS-RFBR programme (project no.
12-02-91055). W. V. P. acknowledges supports from the French
Ministry of Education during his stays in Paris, the Dynasty
Foundation, and RFBR (project no. 12-02-00339).

\end{document}